
\font\lbf=cmbx10 scaled\magstep2

\def\bs{\bigskip}
\def\ms{\medskip}

\def\ni{\noindent}
\def\cl{\centerline}

\def\title#1{\cl{\lbf #1}\ms}
\def\ctitle#1{\bs\cl{\bf #1}\par\nobreak\ms}
\def\stitle#1{\bs{\ni \bf #1}\par\nobreak\ms}

\def\ref#1#2#3#4{#1\ {\it#2\ }{\bf#3\ }#4\par}
\def\ns{\kern-.33333em}
\def\CMP{Communications in Mathematical Physics}

\def\PR{Physical Review}
\def\PRL{Physical Review Letters}

\magnification=\magstep1
\line{\hfill UMD 94-51}
\line{\hfill gr-qc/9310033}
\title{Comments on Dragging Effects}
\ctitle{Dieter R. Brill}
\cl{Department of Physics}
\cl{University of Maryland}
\cl{College Park MD 20742 U.S.A.}
\bs \cl{Submitted to volume in {\it Einstein Studies} series}
\bs \bs \ni {\bf Abstract.} This is a reply, given
at the conference ``Mach's Principle" in T\"ubingen in July 1993,
to the paper by Pfister (1993).
\ms\ni
\stitle{1. Preliminaries and Frivolities}\ni
The ``Machian" effects of rotation on inertial frames have a distinguished
history, as pointed out in the previous paper by Pfister (1993). But beyond
that, the appeal of Mach's ideas has also
been used to motivate points of view that often are
well-founded in inverse relation to their attraction for the general public.
For example,
in a recent ``popular" scientific book (Fahr 1993), Mach occurs
as early as page 55, and there is a whole chapter devoted to matters Machian,
with the somewhat ominous title* {\it The view in the large --- doomed to
failure?} Even
if the answers of this particular book may leave much to be desired, it
reminds us that these questions are not esoteric considerations of a select
few, but of direct and immediate interest to a large number of intellectuals;
and one can only wish that the present volume may have some fraction of
the impact that such ``popular" books have
on the general public!

Pfister's emphasis is not so much on Mach as an abstract principle, but
on definite and calculable effects suggested by the Machian line of thought.
This is a very useful and practical approach;
it is likely that Mach's ideas have brought
more progress by suggesting calculations of such effects than by any
general formulation of his principle. This view has been expressed by
Prof.~Shimony (1992), who compares Mach to a Yankee storekeeper
with many useful items on his shelves, connected to each other only loosely
if at all.
Happily, in Vermont Prof.~Shimony  has actually found  Machs' General Store
(Fig.~1), and characterizes it thus:
``what we have is on the shelves;
what you don't see don't exist; we give no credit."

The dragging effects are probably the most prominent items in
Mach's store, and Pfister gives their interesting history.
One could supplement this with the history of other Machian effects,
because one use of a good principle is to make simple and quick
qualitative or semi-quantitative suggestions for a variety of
different physical situations. The most fruitful and imaginative use
of Mach's principle in this fashion occurred in the 50's and 60's,
largely inspired by Dicke. Some of the resulting ideas
have appeared in the literature (for example, Dicke 1961, 1963, Peebles
and Dicke 1962),
but they found their most creative expression in Dicke's informal
``brainstorming" sessions for the Princeton relativists.
Here all sorts of ways that the universe might influence
local physics were explored, and the magnitudes of possible experimental
observations were determined by liberal use of Mach's principle.

\stitle{2. Dragging of inertial frames}
\ni
Of all the predictions that follow from, or have been read out of, Mach's
principle, the dragging of inertial frames by rotating bodies is certainly
the most definite and least controversial. If one measures this dragging
by the Coriolis forces
--- that is, one calculates in the first order of the angular velocity
--- then the answer of General Relativity is unambiguous, and can be
derived simply and elegantly (Pfister 1993). This answer can be
interpreted to agree with the Machian expectations, at least if one
is willing to refine them suitably and to some extent after the fact.
Mach himself, in his famous passage (Mach 1883), simply informs us
that the dragging might be measured by centrifugal forces, but
it is reasonable to include
Coriolis forces (and other effects!) with respect to Mach's principle.
After all, no one is competent to say how Mach's principle
would have turned out if the background of its author increased in
scope and style till it ultimately encompassed several modern alternative
theories of gravity.

It is the particular merit of Pfister and his students that they did a detailed
study of the Machian centrifugal force terms. By the ``correct"
centrifugal force Pfister means the
one that corresponds to the spacetime region being {\it flat}. As he remarks,
to have a centrifugally correct
region, flat to second order in the angular velocity, necessarily demands a
rather special situation: nonlinearity will generally cause second-order
forces other than pure centrifugal ones. Thus he finds that the correct
forces inside a shell require that the shell have special properties,
such as a prolate shape, differential rotation, etc.

But given these special properties one {\it can}
induce the ``correct" inertial forces for the case of rotation;
in fact, Pfister conjectures that the effects of {\it arbitrary} acceleration
can be induced in a laboratory at rest with respect to the fixed stars
by suitably moving exterior masses. This is a very interesting conjecture,
not least because it forces us to think about the proper definition
of global measurements, such as
acceleration with respect to the fixed stars, and about other global Machian
quantities. For example, naively one might reject the conjecture because the
inertial forces are ``coordinate effects," whereas the effect of exterior
masses is presumably a curvature effect; but this would neglect the global
nature of the acceleration that is at issue here.

In the general situation, when the exterior masses are not moving
``suitably," Pfister shows that it is difficult to distinguish between
Machian dragging and gravitational radiation. If we say that it is in fact
impossible, we come very close to the type of minimalistic view of Mach's
principle advocated for example in the Mach-Einstein-Wheeler formulation
(Isenberg 1993). But just because the Machian dragging is a (global)
``coordinate effect," whereas gravitational radiation is a (local)
``curvature effect"
we can equally well view the difficulty as a challenge, to define the
Machian dragging in a general and useful way so that it {\it can} be
distinguished from gravitational radiation.
\eject

\stitle{3. Alternative Theories}
\ni
This is not the place to review the many ways in which Mach's principle has
suggested alternative theories. In the case of inertial dragging, Mach of
course makes no explicit prediction that might confirm or contradict the
general relativity result; but in the limit when the rotating mass is
the whole universe, the usual expectation is that of ``complete dragging."
It may be that in this limit the small-rotation results will be exact
(because complete dragging presumably means that the rotation with respect
to the inertial frame is vanishingly small).
As a first step toward a proof of such conjectures
nothing would be more helpful than
a clear, general definition of ``complete dragging."

If this cosmological Machian expectation is not satisfied, it may be the
fault of the cosmological model. The task would then be to find a more
restricted class of models that are Machian, as advocated in a number of
formulations (such as the Einstein-Wheeler-Mach version). Or it may be the
fault of the theory, in which case one would look for a more Machian
theory of gravity; this is one of the claims of the Brans-Dicke theory.
Calculations of dragging effects in alternative theories are appropriate
not only to determine this ``degree of Machismo," but also because these
effects will eventually become measurable experimentally, and it is
interesting to know how accurately one will have to measure in order
to distinguish alternative theories on the basis of dragging. This
question has been answered (to lowest order) in the {\it PPN} formalism,
but beyond that the literature appears limited.

Two alternative theories are particularly interesting  in this
context, Brans-Dicke
theory (because it takes Mach's principle as a motivational basis) and
low-energy string theory (because string theory is supposed to be
more fundamental than general relativity). Both of these incorporate
a scalar field (``dilaton") as part of the gravitational force. That
the scalar part of gravity should not contribute to dragging is suggested by
symmetry considerations: to lowest order, rotational perturbations
will affect only
the $g_{0i}$ components of the metric, and {\it not} the one component of
the scalar field. Therefore in Brans-Dicke theory, for example,
the dragging is
reduced compared to general relativity (Brill 1962), because the scalar field
contributes positively to the gravitational force but ``does not drag."
One can also show that this theory is, in a sense, {\it more} Machian
than general relativity: it leads to the
limit of complete dragging in the limit when the rotating shell is
the only matter in
the universe, independent of the shell's mass: the
gravitational ``constant" will automatically adjust so that
the shell is at its Schwarz\-schild radius.

Dragging effects appear not to have been discussed in the literature on
low-energy string theory, which has no natural place for phenomenologically
described matter. The long-range fields in this theory are the gravitational,
scalar (dilaton), electromagnetic, and axion fields.
However, one does have solutions to this theory for rotating black holes
(Sen 1992). For uncharged black holes these show {\it no}
difference from general relativity, and appear to suggest that the dragging is
not reduced as in Brans-Dicke theory. This may be surprising if one
expects the rotating black hole solutions to reproduce, at least to lowest
order, the exterior dragging
effects of other rotating configurations; but the reason is easily
explained.

The black hole solutions of low-energy string theory are the simplest
solutions to the {\it sourceless} equations corresponding to black holes.
In these equations the electromagnetic field generates the scalar field;
if the former vanishes (zero charge), the scalar field is constant.
In the Brans-Dicke theory, on the other hand, one has in mind matter sources,
and the scalar field is generated by the trace of the matter's
stress-energy tensor. Thus the difference already occurs in the
description of an unperturbed mass center: Brans-Dicke theory attributes
a part of the gravitational force from matter to the scalar field,
whereas low-energy string theory considers the scalar field to be constant
for uncharged black holes.
In fact, collapse to a black hole in Brans-Dicke theory also leads
to black holes with a constant scalar field, which are the same
as those of general relativity (Hawking 1972). Thus there is no reason
to suppose that the role of the scalar field in dragging effects is
different in these theories. (Whether there are also other types of
black holes in Brans-Dicke theory is the subject of current research,
Campanelli and Lousto 1993).

\bs\ni
{\bf Footnote}\ms\ni
*Chapter 6, ``Der Blick ins Gro{\ss}e --- zum Scheitern verurteilt?"
\bs\ni
{\bf Figure caption}\ms\ni
Fig.~1: Machs' General Store, Pawlet, Vermont
\bs
\begingroup
\parindent=0pt\everypar={\global\hangindent=20pt\hangafter=1}\par
{\bf References}\ms
\ref{Brans, Carl H. and Dicke, Robert H. ``Mach's Principle and a
Relativistic Theory of Gravitation"}\PR{124}{925-935}
\ref{Campanelli, M. and Lousto, C.~O. ``Are Black Holes in Brans-Dicke
Theory precisely the same as in General Relativity?"}{preprint}\ns
{Universit\"at Konstanz; gr-qc/9301013}
\ref{Dicke, Robert H. 1961 ``Experimental Tests of Mach's Principle"}
\PRL{7}{359-360}
\ref{Dicke, Robert H. 1963 ``Cosmology, Mach's Principle and Relativity"}
{American Journal of Physics}{31}{500-509}
\ref{Dicke, Robert H. 1962 ``The Earth and Cosmology"}
{Science}{138}{653-664}
\ref{Fahr, Hans J\"org 1993}{Der Urknall kommt zum Fall}{}{Frankh-Kosmos}
\ref{Hawking, K.~Stephen 1972 ``Black Holes in the Brans-Dicke Theory
of Gravitation"}\CMP{25}{167-171}
\ref{Isenberg, James 1993 ``On the Wheeler-Einstein-Mach Spacetimes"}{}{}
{this volume}
\ref{Peebles, P. James and Dicke, Robert H. 1962 ``Significance of Spatial
Isotropy"}\PR{127}{629-631}
\ref{Pfister, Herbert 1993 ``Dragging effects near rotating bodies and in
cosmological models"}{}{}{previous paper, this volume}
\ref{Sen, A. 1992 ``Rotating Charged Black Hole Solutions in Heterotic String
Theory"}\PRL{69}{1006}
\ref{Shimony, Abner 1992}{private communication}\ns{APS Spring Meeting,
Washington, DC}
\endgroup
\end